\documentclass[reqno]{amsart}

\usepackage{booktabs} 
\usepackage{IEEEtrantools}
\usepackage{amssymb,latexsym,amsfonts,amsmath}
\usepackage{graphicx}
\usepackage{mathrsfs}
\usepackage{dsfont}

\usepackage{subfigure}

\usepackage{tikz}
\usetikzlibrary{calc,shapes,arrows}
\usepackage{subfig}

\usepackage{algorithm}
\usepackage{algorithmic}

\usepackage{xspace}

\newtheorem{theorem}{Theorem}[section]
\newtheorem{lemma}[theorem]{Lemma}

\newtheorem{problem}[theorem]{Problem}

\newtheorem{definition}[theorem]{Definition}

\newtheorem{remark}[theorem]{Remark}
\newtheorem{assumption}{Assumption}
\numberwithin{equation}{section}

\newcommand{\R}{{\mathbb{R}}}

\newcommand{\N}{{\mathbb{N}}}

\newcommand{\Let}{:=}
\newcommand{\EE}{\mathds{E}}
\newcommand{\PP}{\mathds{P}}

\usepackage[many]{tcolorbox}
\usetikzlibrary{calc}
\tcbuselibrary{skins}

\newtcolorbox{resp}[1][]{%
	enhanced jigsaw,%
	colback=gray!5!white,%
	colframe=gray!80!black,%
	size=small,%
	boxrule=1pt,%
	halign title=flush center,%
	coltitle=black,%
	breakable,%
	drop shadow=black!50!white,%
	attach boxed title to top left={xshift=1cm,yshift=-\tcboxedtitleheight/2,yshifttext=-\tcboxedtitleheight/2},%
	minipage boxed title=3cm,%
	boxed title style={%
		colback=white,%
		size=fbox,%
		boxrule=1pt,%
		boxsep=2pt,%
		underlay={%
			\coordinate (dotA) at ($(interior.west) + (-0.5pt,0)$);
			\coordinate (dotB) at ($(interior.east) + (0.5pt,0)$);
			\begin{scope}[gray!80!black]
				\fill (dotA) circle (2pt);
				\fill (dotB) circle (2pt);
			\end{scope}
		}%
	},%
	#1%
}

\usepackage{fancyhdr}

\newenvironment{nouppercase}{%
	\renewcommand{\uppercasenonmath}[1]{}}{}

\begin{document}

\begin{abstract}
This paper is concerned with a data-driven technique for constructing finite Markov decision processes (MDPs) as finite abstractions of discrete-time stochastic control systems with \emph{unknown} dynamics while providing formal closeness guarantees. The proposed scheme is based on notions of \emph{stochastic bisimulation functions} (SBF) to capture the probabilistic distance between state trajectories of an unknown stochastic system and those of finite MDP. In our proposed setting, we first reformulate corresponding conditions of SBF as a robust convex program (RCP). We then propose a scenario convex program (SCP) associated to the original RCP by collecting a finite number of data from trajectories of the system. We ultimately construct an SBF between the data-driven finite MDP and the unknown stochastic system with a given confidence level by establishing a probabilistic relation between optimal values of the SCP and the RCP. We also propose two different approaches for the construction of finite MDPs from data. We illustrate the efficacy of our results over a \emph{nonlinear} jet engine compressor with unknown dynamics. We construct a data-driven finite MDP as a suitable substitute of the original system to synthesize controllers maintaining the system in a safe set with some probability of satisfaction and a desirable confidence level.
\end{abstract}

\title{{\Large Constructing MDP Abstractions Using Data with Formal Guarantees}$^*$\footnote[1]{$^*$This work is supported in part by the Swiss National Science Foundation under NCCR Automation (grant agreement 51NF40-180545) and by the NSF under grant CNS-2145184.}}

\author{{\bf {\large Abolfazl Lavaei}}$^1$}
\author{{\bf {\large Sadegh Soudjani}}$^2$}
\author{{\bf {\large Emilio Frazzoli}}$^1$}
\author{{\bf {\large Majid Zamani}$^3$}\\
{\normalfont $^1$Institute for Dynamic Systems and Control, ETH Zurich, Switzerland}\\
{\normalfont $^2$School of Computing, Newcastle University, United Kingdom}\\
{\normalfont $^3$Computer Science Department, University of Colorado Boulder, USA, and, LMU Munich, Germany}\\
\texttt{\{alavaei,efrazzoli\}@ethz.ch}, \texttt{sadegh.soudjani@ncl.ac.uk}, \texttt{majid.zamani@colorado.edu}}

\pagestyle{fancy}
\lhead{}
\rhead{}
\fancyhead[OL]{Abolfazl Lavaei, Sadegh Soudjani, Emilio Frazzoli, and Majid Zamani}

\fancyhead[EL]{Constructing MDP Abstractions Using Data with Formal Guarantees} 
\rhead{\thepage}
\cfoot{}

\begin{nouppercase}
	\maketitle
\end{nouppercase}

\section{Introduction}

Formal controller synthesis for stochastic systems is very challenging mainly due to (i) stochastic nature of dynamics and (ii) the computational complexity arising from uncountable sets of states of underlying systems, especially in real-life safety-critical
applications. To alleviate the encountered complexity, one promising approach is to employ finite abstractions of given systems as suitable replacements in the controller synthesis procedures. In this respect, one can first abstract the original system by a simpler one with finite-state sets (\emph{a.k.a.,} finite Markov decision process (MDP)), perform analysis and synthesis over the abstract model, and finally carry the results back over the concrete system. Since the probabilistic distance between original systems and their finite MDPs lives within a guaranteed error bound, one can ensure that original systems also fulfill the same property as finite MDPs with a quantified probabilistic error.

In the past two decades, there have been many studies on the abstraction-based analysis of stochastic systems. Existing results include finite abstractions for safety and reachability of stochastic systems~\cite{APLS08}; approximation techniques for jump-diffusion systems~\cite{julius2009approximations}; finite abstractions based on adaptive and sequential gridding~\cite{SA13}; finite bisimilar abstractions for incrementally stable stochastic switched systems~\cite{zamani2015symbolic} and stochastic control systems~\cite{zamani2014symbolic}; and finite abstractions for mapping a discrete-time stochastic system to an interval Markov chain~\cite{lahijanian2015formal}, to name a few. \emph{Compositional} abstraction-based techniques have been also proposed to deal with the underlying \emph{curse of dimensionality} as the main bottleneck in the construction of (in)finite abstractions. Existing results include compositional finite abstractions for stochastic control~\cite{lavaei2018ADHSJ,lavaei2017HSCC,lavaei2019NAHS}, switched~\cite{lavaei2019HSCC_J,lavaei2022LSS_J}, and hybrid systems~\cite{hahn2013compositional}, compositional finite MDPs based on dynamic Bayesian networks~\cite{SAM15}, 
and compositional infinite abstractions for stochastic control systems~\cite{lavaei2017compositional,lavaei2018CDCJ}, to name a few. 

Unfortunately, all the above-mentioned literature on the construction of finite abstractions are model-based.
On the other hand, obtaining an accurate model via \emph{identification techniques}, as a potential solution proposed in the relevant literature, is always challenging, time-consuming, and expensive, especially if the underlying dynamics are too complex which is the case in many real-world applications (see \emph{e.g.,}~\cite[and references herein]{Hou2013model}). Hence, there is a need to bypass the system identification phase and directly construct finite MDPs by collecting data from trajectories of the system.

\begin{figure}[h!]
	\centering 
	\includegraphics[width=0.7\linewidth]{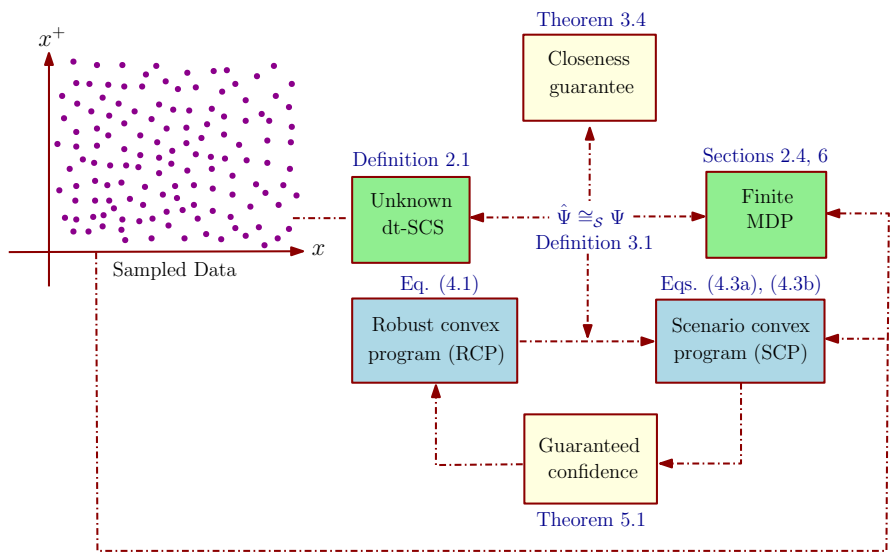}
	\caption{A graphical representation of the  structure of the paper and its contributions.}
	\label{Diagram}
\end{figure}

The main contribution of this work is to develop a data-driven approach for constructing finite MDPs for stochastic control systems with unknown dynamics. A graphical
	representation of the structure of the paper and its contributions are illustrated in Fig.~\ref{Diagram}. As it can be observed, we quantify the probabilistic distance between original systems and their data-driven finite MDPs via a notion of so-called \emph{stochastic bisimulation functions} (SBF). To do so, we first cast the required conditions of SBF as a robust convex program (RCP) and then propose a scenario convex program (SCP) corresponding to the original RCP by collecting a finite number of data from trajectories of the system. We then relate optimal values of the SCP and the RCP, and consequently, construct an SBF between the unknown system and its data-driven finite MDP based on the number of data and an a-priori confidence level. Note that although the proposed results in~\cite{campi2008exact,campi2021scenario} relate  SCP and chance-constrained programs (CCP), we transfer guarantees all the way from SCP to the original RCP which is the main problem in our setting. We show the effectiveness of our proposed data-driven approaches over a \emph{nonlinear} jet engine compressor with unknown dynamics.
    We provide proofs of all statements in Appendix.

Data-driven constructions of finite abstractions via a Gaussian process regression and a probably approximately correct (PAC) statistical approach are, respectively, proposed in~\cite{hashimoto2020learning,devonport2021symbolic}. While the results in~\cite{hashimoto2020learning,devonport2021symbolic} are for the construction of \emph{finite transition systems} with unknown dynamics, we develop here a data-driven approach for constructing \emph{finite MDPs} which is more challenging. In addition, the proposed approach in~\cite{hashimoto2020learning} assumes that the underlying system consists of both known and unknown dynamics and the main goal is to learn unknown dynamics via Gaussian processes. In comparison, we propose a \emph{direct data-driven} approach to construct finite MDPs from unknown models without performing any system identification. 
Data-driven safety verification of stochastic systems with unknown dynamics is proposed in~\cite{Ali_ADHS21} but using barrier certificates (if existing) rather than constructing finite MDPs (always existing) which is the case in our work. 
A statistical model checking algorithm for MDPs and stochastic games with PAC guarantees is proposed in~\cite{ashok2019pac} by assuming the knowledge of either the underlying graph or a minimum bound on transition probabilities. A scenario approach to compute PAC probability intervals using model-based abstractions is presented in~\cite{badings2021sampling} with the focus on linear systems. In comparison, we propose data-driven approaches for the construction of (interval) MDPs for any class of \emph{nonlinear} systems without requiring any knowledge from underlying systems. 

\section{Discrete-Time Stochastic Control Systems}\label{Sec: dt-NDS}

\subsection{Notation}

Sets of real, positive and non-negative real numbers are denoted by $\mathbb{R},\mathbb{R}^+$, and $\mathbb{R}^+_0$, respectively. We denote the sets of non-negative and positive integers by $\mathbb{N} := \{0,1,2,...\}$ and $\mathbb{N}^+=\{1,2,...\}$, respectively. The power set of any set $X$, as the set of all subsets of $X$, is denoted by $2^X$. Given $N$ vectors $x_i \in \mathbb{R}^{n_i}$, $x=[x_1;...;x_N]$ denotes the corresponding column vector of dimension $\sum_i n_i$. Minimum and maximum eigenvalues of a symmetric matrix $A$ are denoted by $\lambda_{\min}(A)$ and $\lambda_{\max}(A)$, respectively. The absolute value of $a\in\mathbb R$ is denoted by $\vert a\vert$. We denote the Euclidean norm of a vector $x\in\mathbb{R}^{n}$ by $\Vert x\Vert$. For any matrix $P\in\mathbb R^{m\times n}$, we have $\|P\| := \sqrt{\lambda_{\max}(P^\top P)}$. The (essential) supremum of a measurable function $f:\mathbb N\rightarrow\mathbb{R}^n$ is denoted by $\Vert f\Vert_{\infty} \Let \text{(ess)sup}\{\Vert f(k)\Vert,k\geq 0\}$. If a system $\Psi$ fulfills a property $\varphi$, it is denoted by $\Psi \vDash \varphi$. 
The operator $\vDash$ is also employed to show the feasibility of a solution for an optimization problem. A function $\varphi: \mathbb{R}^+_0 \rightarrow \mathbb{R}^+_0$ is said to be a class $\mathcal{K}$ function if it is continuous, strictly increasing, and $\varphi(0)=0$. A class $\mathcal{K}$ function $\varphi(\cdot)$ is called a class $\mathcal{K}_\infty$ if $\varphi(s) \rightarrow \infty$ as $s \rightarrow \infty$.

\subsection{Preliminaries}
We define a probability space as $(\Omega, \mathcal{F}_\Omega, \PP_\Omega)$, where $\Omega$ is the sample space, $\mathcal{F}_\Omega$ is a sigma-algebra on $\Omega$ as events, and $\PP_\Omega$ is the probability measure. Given the probability space $(\Omega,\mathcal F_{\Omega},\mathds{P}_{\Omega})$, we denote the $N$-Cartesian product set of $\Omega$ by $\Omega^N$, and its corresponding product measure by $\PP^N$. 
Random variables are assumed to be measurable functions of the form $X:(\Omega,\mathcal{F}_\Omega) \rightarrow (S_X,\mathcal{F}_X)$ so that $X$ induces a probability measure on $(S_X,\mathcal{F}_X)$ as $Prob\{A\} = \mathbb{P}_\Omega\{X^{-1}(A)\}$ for any $A \in \mathcal{F}_X.$
A random variable $\varsigma$ with standard normal distributions (zero mean and variance of identity) is denoted by $\varsigma(\cdot) \sim\mathcal N(0, \mathds{I}_n)$. The topological space $S$ is a Borel space if it is homeomorphic to a Borel subset of a Polish space, \textit{i.e.}, a separable and completely metrizable space. 
The Borel sigma-algebra from Borel space $S$ is denoted by $\mathcal{B}(S)$. The map $f: S \rightarrow Y$ is measurable whenever it is Borel measurable.

\subsection{Discrete-Time Stochastic Control Systems}\label{Sec: dt-SCS}

Here, we consider discrete-time stochastic control systems (dt-SCS) as in the following definition.

\begin{definition}
	A discrete-time stochastic control system (dt-SCS) is characterized by the tuple
	\begin{align}\label{EQ:13}
		\Psi=(X,U,\varsigma,f),
	\end{align}
	where:
	\begin{itemize}
		\item $X\subseteq \mathbb R^n$ is a Borel set as the state set of the system;
		\item $U = \{u_1,u_2,\dots,u_m\}\subseteq \mathbb R^{\bar m}$ is the finite input set of the system;
		\item $\varsigma$ is a sequence of independent-and-identically distributed
		(i.i.d.) random variables from a sample space $\Omega$ to a set
		$\mathcal{V}_{\varsigma}$, \emph{i.e.,}	$\varsigma:=\{\varsigma(k):\Omega\rightarrow \mathcal V_{\varsigma},\,\,k\in\N\}$;
		\item $f:X\times U\times \mathcal{V}_{\varsigma}\rightarrow X$ is a measurable function describing the state evolution of the system.
	\end{itemize}
\end{definition}
The transition map $f$  and distribution of $\varsigma$ are both assumed to be \emph{unknown} in this work. 

For a given initial state $x(0)\in X$ and an input sequence $\nu(\cdot):\mathbb N\rightarrow U$, evolution of the state of dt-SCS $\Psi$  is characterized by

\begin{align}\label{EQ:14}
	\Psi:x(k+1)=f(x(k),\nu(k),\varsigma(k)),\quad k\in\mathbb N. 
\end{align} 
The random sequence $x_{a\nu}: \Omega \times \mathbb N \rightarrow X $ denotes the state trajectory of $\Psi$ at time $k\in\mathbb N$ under an input sequence $\nu(\cdot)$ starting from an initial condition $x(0)= a$.

\subsection{Finite Markov Decision Processes}\label{MDPs}

A dt-SCS $\Psi$ in \eqref{EQ:13} can be \emph{equivalently} characterized as a \emph{continuous-space} MDP~\cite{kallenberg1997foundations}
\begin{equation}\notag
	\Psi=(X,U,\mathsf{T}_{\mathsf x}),	
\end{equation}
where the map $\mathsf{T}_{\mathsf x}:\mathcal B(X)\times X\times U\rightarrow[0,1]$ is a conditional stochastic kernel that associates to any $x \in X$, a probability measure $\mathsf{T}_{\mathsf x}(\cdot \,\big |\, x,u)$
on
$(X,\mathcal B(X))$
so that for any set $\mathcal A \in \mathcal B(X)$,
$$\PP \big\{x(k+1)\in \mathcal A\,\big |\, x(k),\nu(k)\big\} = \int_{\mathcal A} \mathsf{T}_{\mathsf x} (\mathsf{d}x(k+1)\,\big |\,x(k),\nu(k)).$$
The conditional stochastic kernel $\mathsf{T}_{\mathsf x}$ can be \emph{uniquely} determined by the pair $(\varsigma,f)$~\cite{kallenberg1997foundations}. 
\begin{assumption}\label{compact}
	Here, we assume that the state set $X$ is compact which is the case in many practical applications. The rest of the states outside the compact region can be considered as a single absorbing state.
\end{assumption}
Under Assumption~\ref{compact}, we first construct finite partitions of the state set as $X = \cup_i \mathsf X_i$
	and then select representative points $\bar x_i\in \mathsf X_i$ as abstract states. Given a continuous-space MDP $\Psi=(X,U,\mathsf{T}_{\mathsf x})$, its \emph{finite} MDP can be constructed by $\hat \Psi=(\hat X, U,\mathsf{\hat T}_{\mathsf x})$, where the discrete transition probability matrix $\mathsf{\hat T}_{\mathsf x}$ is defined as
\begin{align*}
	\mathsf{\hat T}_{\mathsf x} (x'\,\big|\,x,u) 
	= \mathsf{T}_{\mathsf x} (\Xi(x')\,\big|\,x,u), ~\forall x,x' \in \hat X, ~\forall u\in U,
\end{align*}
with the map $\Xi:X\rightarrow 2^X$ that assigns to any $x\in X$, the corresponding partition set it belongs to, \emph{i.e.,} $\Xi(x) = \mathsf X_i$ if $x\in \mathsf X_i$ for some $i=1,2,\ldots,n_{\bar x}$.
Equivalently, given a dt-SCS $\Psi=(X,U,\varsigma,f)$, its \emph{finite MDP} can be constructed as
\begin{equation*}
	\hat\Psi =(\hat X, U,\varsigma,\hat f),
\end{equation*}
where $\hat X := \{\bar x_i, i=1,...,n_{\bar x}\}$ is the finite state set of~$\hat\Psi$. Moreover, $\hat f:\hat X\times U \times\mathcal V_{\varsigma}\rightarrow\hat X$ is defined as
\begin{equation}\label{EQ:3}
	\hat f(\hat{x},u,\varsigma) = \Pi_{x}(f(\hat{x},u,\varsigma)).
\end{equation}
The map $\Pi_x:X\rightarrow \hat X$ assigns to any $x\in X$, a representative point $\bar x\in\hat X$ of the corresponding partition set and fulfills the inequality
\begin{equation}\label{EQ:4}
	\Vert \Pi_x(x)-x\Vert \leq \eta,\quad \forall x\in X, 
\end{equation}
where $\eta:=\sup\{\|x-x'\|,\,\, x,x'\in \mathsf X_i,\,i=1,2,\ldots,n_{\bar x}\}$ is the \emph{state discretization parameter}.
\begin{remark}
	Note that for constructing finite MDPs, we do not put any constraints on the shape of partition sets. However, for the sake of an easier implementation, we consider partition sets as boxes (hyper-rectangle in high dimensions) and center of each box as representative points.
\end{remark}

\section{Stochastic Bisimulation Functions}
Here, we present a notion of so-called \emph{stochastic bisimulation functions} to quantify the probabilistic closeness between state trajectories of the dt-SCS $\Psi$ and its
finite MDP $\hat\Psi$.

\begin{definition}\label{Def:3}
	Consider a dt-SCS $\Psi =(X,U,\varsigma,f)$ and its 
	finite MDP $\hat\Psi =(\hat X,U,\varsigma,\hat f)$. A function $\mathcal S:X\times\hat X\to\R_0^+$ is called a stochastic bisimulation function (SBF) between $\hat\Psi$ and $\Psi$, denoted by $\hat\Psi\cong_{\mathcal{S}}\Psi$, if there exist
	$\alpha\in\R^+$ and $\psi \in\R_0^+$ such that
	for all $x\in X$, $\hat x\in\hat X$, $u\in U$, one has
	\begin{align}\label{EQ:15}
		\alpha\Vert x - \hat x\Vert^2\leq \mathcal S(x,\hat x)
	\end{align}
	\begin{align}\label{EQ:16}
		&\EE \Big [\mathcal S(f(x(k),\nu(k),\varsigma(k)),\hat{f}(\hat x(k),\nu(k),\varsigma(k)))\big | x(k), \hat x(k), \nu(k)\Big ]\leq\mathcal S(x(k),\hat{x}(k))+\psi,
	\end{align}
	where the expected value $\EE$ is with respect to $\varsigma$ under the one-step transition of dt-SCS and its finite MDP.
\end{definition}
\begin{remark}
	Stochastic bisimulation
	functions in Definition~\ref{Def:3} roughly speaking guarantee that if the concrete system and its abstraction start from two close initial conditions (expressed by condition~\eqref{EQ:15}), then their states remain close in terms of expectation after a one-step transition (expressed by condition~\eqref{EQ:16}). In particular, condition~\eqref{EQ:16} ensures that $\mathcal{S}$ is a $c$-martingale process~\cite{1967stochastic} ensuring the probability bound in Theorem~\ref{Thm:4}. These types of conditions are stochastic counterparts of the ones in the notions of (bi)simulation relations~\cite{tabuada2009verification}.
\end{remark}
\begin{remark}
		Note that for the sake of an easier presentation, we assumed that $\alpha$ in~\eqref{EQ:15} is linear in $\Vert x - \hat x\Vert^2$. However, one can readily extend it to be a $\mathcal{K}_{\infty}$ function but at the cost of introducing extra decision variables in RCP~\eqref{RCP}, and accordingly, solving $\text{SCP}_\varsigma$~\eqref{SCP1} with more number of data (cf. Theorem~\ref{Thm:6}).
\end{remark}
In the next theorem, borrowed from~\cite{lavaei2018CDCJ}, we employ SBF in Definition~\ref{Def:3} and quantify the probabilistic distance between state trajectories of $\Psi$ and those of its finite MDP $\hat\Psi$ within finite time horizons.

\begin{theorem}\label{Thm:4}
	Consider a dt-SCS $\Psi =(X,U,\varsigma,f)$ and its 
	finite MDP $\hat\Psi =(\hat X,U,\varsigma,\hat f)$. Suppose $\mathcal S$ is an SBF between $\hat\Psi$ and $\Psi$ as in Definition~\ref{Def:3}. Then for any random variables $a$ and $\hat a$ as initial states of $\Psi$ and $\hat\Psi$, the probabilistic distance between state trajectories of dt-SCS and those of its
	finite MDP can be quantified within a time horizon $\mathcal T\in \mathbb N^+$ as
	\begin{align*}
		&\PP\left\{\sup_{0\leq k\leq \mathcal T}\Vert x_{a\nu}(k)-\hat x_{\hat a\nu}(k)\Vert\geq\varepsilon\,\big |\,a,\hat a\right\}\leq \frac{\mathcal S(a,\hat a) + \psi \mathcal T}{\alpha\varepsilon^2},
	\end{align*}
\end{theorem}
where $\varepsilon\in \mathbb R^+$ is chosen arbitrarily, and $x_{a\nu}(k)$ and $\hat x_{a\nu}(k)$ denote, respectively, state trajectories of $\Psi$ and $\hat \Psi$ at time $k\in\mathbb N$ under an input sequence $\nu(\cdot)$ starting from initial conditions $x(0)= a$ and $\hat x(0)= \hat a$. If $\psi=0$ in \eqref{EQ:16}, the time horizon $\mathcal T$ can be chosen arbitrarily large without affecting the probability upper bound in Theorem \ref{Thm:4}.

\begin{remark}
	As shown in~\cite[Section 6]{lavaei2018CDCJ}, given a \emph{finite-time horizon} property $\varphi$, one can utilize the results of Theorem~\ref{Thm:4} and quantify a lower bound for $\PP\big\{\Psi\vDash\varphi\big\}$ via $\PP\big\{\hat\Psi\vDash\varphi^\varepsilon\big\}$ as
	\begin{align}\label{epsilon-perturbed}
		\PP\big\{\hat\Psi\vDash\varphi^\varepsilon\big\} - \delta \le \PP\big\{\Psi\vDash\varphi\big\},
	\end{align}
	where $\delta = \frac{\mathcal S(a,\hat a) + \psi \mathcal T}{\alpha\varepsilon^2}$ and $\varphi^\varepsilon$ is an $\varepsilon$-deflated version of $\varphi$ defined as~\cite[Section 6]{lavaei2018CDCJ}
	\begin{equation*}
		\varphi^\varepsilon \Let \{x\in\varphi\,\,\big|\,\,\| x' - x\|\geq\varepsilon\, \text{ for all }x'\in X\backslash\varphi\}.
	\end{equation*}
\end{remark}
We now present the main problem that we aim to solve.

\begin{resp}
	\begin{problem}\label{Prob:2}
		Consider a dt-SCS $\Psi$ in~\eqref{EQ:14} with unknown $f$ and distribution of $\varsigma$. Develop a data-driven approach for the construction of an SBF between $\hat\Psi$ and $\Psi$ with an a-priori confidence bound $1-\beta \in(0,1)$ as
		\begin{align*}
			\PP^N\big\{\hat\Psi\cong_{\mathcal{S}}\Psi\big\}\ge 1-\beta.
		\end{align*}
	\end{problem}
\end{resp}

\section{Data-Driven Construction of SBF}
Here, we fix the structure of SBF as $\mathcal{S}(q,x,\hat x)=\sum_{j=1}^{r} {q}_jb_j(x,\hat x)$ with some user-defined (possibly nonlinear) basis functions $b_j(x,\hat x)$ and unknown coefficients $q=[{q}_{1};\ldots;q_r] \in \mathbb{R}^r$. In order to enforce the required conditions for the construction of SBF as in~\eqref{EQ:15}-\eqref{EQ:16}, we reformulate the problem as the following robust convex program (RCP) with an optimal value denoted by $\Upsilon_R^*$:
\begin{align}\label{RCP}
	&\text{RCP}\!:\!\left\{
	\hspace{-1.5mm}\begin{array}{l}\min\limits_{[\Phi;\Upsilon]} \quad\!\!\!\Upsilon,\\
		\, \text{s.t.} \quad  \,\max_j\big\{g_j(x,\hat x,u, \Phi)\big\}\leq \Upsilon,~  j\in\{1,2\}, \\ 
		\quad\quad\quad\!\forall x\in X, ~\forall \hat x\in \hat X, ~\forall u\in U,\\
		\quad\quad\quad\!\Phi = [\alpha;\psi;{q}_{1};\dots;q_r],\alpha\in\R^+,\psi \in\R_0^+,\Upsilon \in \mathbb R,\end{array}\right.
\end{align}
where:
\begin{align}\notag
	g_1(x,\hat x,u,\Phi)& = \alpha\Vert x - \hat x\Vert^2 - \mathcal S(q,x,\hat x),\\\label{EQ:11}
	g_2(x,\hat x,u,\Phi)& = \EE \Big [\mathcal S(q,f(x,u,\varsigma),\hat{f}(\hat x,u,\varsigma))\,\big |\, x, \hat x, u\Big ]-\mathcal S(q,x,\hat{x})-\psi.
\end{align}
If $\Upsilon_R^* \leq 0$, a solution to the RCP implies the satisfaction of conditions~\eqref{EQ:15}-\eqref{EQ:16} for the construction of SBF. 
\begin{remark}
	Note that the RCP in~\eqref{RCP} is convex with respect to \emph{decision variables}. In particular, since the SBF is defined as the linear combination of basis functions (\emph{i.e.,} $\mathcal{S}(q,x,\hat x)=\sum_{j=1}^{r} {q}_jb_j(x,\hat x)$), the RCP in~\eqref{RCP} remains always convex with respect to decision variables.
\end{remark}
As it can be observed, solving the proposed RCP in~\eqref{RCP} is not tractable in general since maps $f$, $\hat f$ and distribution of $\varsigma$ appearing in $g_2$ are all unknown. To tackle this problem, we propose a scenario convex program (SCP) associated to the original RCP by collecting a finite number of data from trajectories of the system. Let $\{x_i\}^{N}_{i=1}$ denote $N$ i.i.d. data sampled within $X$. Instead of solving the RCP in~\eqref{RCP}, we provide the following scenario convex program (SCP) with an optimal value denoted by $\Upsilon_N^*$:
\begin{subequations}
	\begin{align}\label{SCP}
		&\text{SCP}_N\!:\!\left\{
		\hspace{-1.5mm}\begin{array}{l}\min\limits_{[\Phi;\Upsilon]} \,\,\,\,\,\!\!\Upsilon,\\
			\, \text{s.t.}  \quad \,\max_j\big\{g_j(x_i, \hat x,u,\Phi)\big\}\leq \Upsilon, ~ j\in\{1,2\}, \\
			\quad \quad\quad \!\!\forall x_i\in X, \forall i\in \{1,\ldots,N\}, \forall \hat x\in \hat X,\forall u\in U,\\
			\quad\quad\quad\!\!\Phi = [\alpha;\psi;{q}_{1};\dots;q_r],\alpha\in\R^+,\psi \in\R_0^+,\Upsilon \in \mathbb R,\end{array}\right.
	\end{align}
	where $g_1$,$g_2$ are the same functions as defined in~\eqref{EQ:11}. Note that $f(x_i,u,\varsigma)$ in $g_2$ can be acquired by measuring the state of unknown dt-SCS after one-step transition starting from $x_i$ and under input $u$. As for computing $\hat f(\hat x,u,\varsigma)$ in $g_2$, one needs to initialize the system at $\hat x$ and feed input $u$ to obtain $f(\hat x,u,\varsigma)$. Given a state discretization parameter $\eta$, $\hat f(\hat x,u,\varsigma)$ is then obtained as the \emph{nearest representative point} to the value of $f(\hat x,u,\varsigma)$, where condition~\eqref{EQ:4} holds.
	
	Although the issue of unknown maps $f,\hat f$ got resolved, the closed-form solution for the expected value in $g_2$ with respect to $\varsigma$ is still needed. To tackle this problem, we employ an empirical approximation of the expected value and propose a new version of SCP, denoted by SCP$_\varsigma$, with an optimal value denoted by $\Upsilon_\varsigma^*$ as:
	\begin{align}\label{SCP1}
		&\text{SCP}_\varsigma\!:\!\left\{
		\hspace{-1.5mm}\begin{array}{l}\min\limits_{[\Phi;\Upsilon]} \,\,\,\,\,\!\!\Upsilon,\\
			\, \text{s.t.}  \quad \,\max\big\{g_{1}(x_i, \hat x,u,\Phi), \bar g_{2}(x_i,\hat x,u,\Phi)\big\}\leq \Upsilon, \\
			\quad \quad\quad\!\forall x_i\in X, \forall i\in \{1,\ldots,N\}, \forall \hat x\in \hat X,\forall u\in U,\\
			\quad\quad\quad\!\Phi = [\alpha;\psi;{q}_{1};\dots;q_r],\alpha\in\R^+,\psi \in\R_0^+,\Upsilon\in \mathbb R,\end{array}\right.
	\end{align}
\end{subequations}
with
\begin{align}\notag
	\bar{g}_{2}(x_i,\hat{x},u,\Phi)=\frac{1}{M}\sum_{z=1}^{M}\mathcal{S}(q,f(x_i,u,\varsigma_{z}),\hat f(\hat x,u,\varsigma_{z}))- \mathcal{S}(q,x_i,\hat{x})-\psi + \mu,
\end{align}
where $M\in\mathbb{N}_{\geq 1}$ and $\mu\in \mathbb{R}_0^+$ are, respectively, the number of noise realizations required for the empirical approximation and the associated error introduced by this approximation.

In the next lemma, we employ Chebyshev's inequality~\cite{saw1984chebyshev} to establish a relation between solutions of SCP$_\varsigma$  and SCP$_N$  with the required number of noise realizations $M$, an a-priori approximation error $\mu$, and a desired confidence $\beta_{1}\in (0,1]$. 

\begin{lemma}\label{Lem:2}
	Suppose $\bar{\mathcal{S}}(q,x,\hat x)$ is a feasible solution for SCP$_\varsigma$ in~\eqref{SCP1}. For an a-priori approximation error $\mu\in \mathbb{R}_0^+$, a desired confidence $\beta_{1}\in (0,1]$, and an upper bound $\mathcal{Q}\in \mathbb{R}^+$ on the variance of SBF applied on $f, \hat f$, \emph{i.e.,} $\text{Var}\big[\mathcal S(q,f(x,u,\varsigma),\hat{f}(\hat x,u,\varsigma))\big]\leq \mathcal{Q}, \forall x \in X, \forall \hat x \in \hat X, \forall u \in U$, one has
	\begin{align*}
		\PP\Big\{\bar{\mathcal{S}}(q,x,\hat x) \models\text{SCP}_N\Big\}\geq 1- \beta_{1},
	\end{align*}
	provided that the number of samples in the empirical approximation
	fulfills $M \geq \frac{\mathcal{Q}}{\beta_{1}\mu^2}$.
\end{lemma}

\section{Out-of-Sample Performance Guarantees}

Here, inspired by the fundamental results of~\cite{esfahani2014performance}, we establish a formal relation between optimal values of $\text{SCP}_\varsigma$ and $\text{RCP}$, and accordingly, formally construct an SBF between $\hat\Psi$ and $\Psi$ based on the number of data and a required confidence level. 

\begin{theorem}\label{Thm:6}
	Consider an unknown dt-SCS as in~\eqref{EQ:14}. Let $g_1, g_2$ be Lipschitz continuous with respect to $x$ with Lipschitz constants $\mathscr{L}_{g_1}, \mathscr{L}_{g_2}$, respectively. Consider the $\text{SCP}_\varsigma$ in~\eqref{SCP1} with its associated optimal value $\Upsilon^*_\varsigma$ and solution $\Phi^* = [\alpha^*;\psi^*;{q}^*_{1};\dots;q^*_r]$, with $N\geq \bar N\big(\varepsilon_{2},\beta_2\big)$, where
	\begin{align}\label{EQ:12}
		\bar N(\varepsilon_{2},\beta_2):=
		\min\Big\{N\in\N \,\big|\,\sum_{i=0}^{c-1}\binom{N}{i}\varepsilon_2^i(1-\varepsilon_2)^{N-i}\leq\beta_2\Big\},
	\end{align}
	$\beta_2 \in [0,1]$, $\varepsilon_{2}=(\frac{\varepsilon_{1}}{\mathscr{L}_{g}})^{n}$, with $\varepsilon_{1}\in [0,1] \leq \mathscr{L}_{g} := \max \big\{\mathscr{L}_{g_1},\mathscr{L}_{g_2}\big\}$,  and with $n,c$ being, respectively, dimension of the state set, and number of decision variables in $\text{SCP}_\varsigma$. If $$\Upsilon^*_\varsigma + \varepsilon_{1} \leq 0,$$ then the constructed $\mathcal S$ from $\text{SCP}_\varsigma$ in~\eqref{SCP1} is an SBF between $\hat\Psi$ and $\Psi$  with a confidence of at least $1-\beta$ with $ \beta = \beta_1 + \beta_2$, \emph{i.e.,}
	\begin{align*}
		\PP^N\big\{\hat\Psi\cong_{\mathcal{S}}\Psi\big\}\ge 1-\beta_1 - \beta_2,
	\end{align*}
	where $\beta_1\in(0,1]$ is a desired confidence for the empirical approximation as presented in Lemma~\ref{Lem:2}. 
\end{theorem}
\begin{remark}
	Note that the SCP$_\varsigma$ in~\eqref{SCP1} may be computationally intractable for large-scale systems given that the minimum number of data in~\eqref{EQ:12} required for solving SCP~\eqref{SCP} is exponential with respect to the dimension of unknown systems.
\end{remark}

In order to compute the required number of data in Theorem~\ref{Thm:6}, one needs to first compute $\mathscr{L}_{g}$. We propose an explicit way to compute $\mathscr{L}_{g}$ for both linear and nonlinear discrete-time stochastic systems in Appendix.

\section{Data-Driven Construction of Finite MDPs}
In this section, we propose two different approaches for the construction of finite MDPs from data.
The constructed finite MDPs can be utilized as suitable substitutes of original systems to perform verification and synthesis over unknown original stochastic systems. The constructed SBF from data can be also leveraged to capture the probabilistic distance between state trajectories of unknown stochastic systems and their data-driven finite MDPs.

\subsection{Interval MDPs from Data}
Here, we first propose an approach to construct interval Markov decision processes (IMDPs) from data and then utilize the constructed IMDPs to transfer our guarantees over finite MDPs. Consider the following Chernoff bound~\cite{chernoff1952measure}:
\begin{align}\label{Chernoff}
	\hspace{0.55cm}\PP\Big\{\vert p_{ij}-\bar p_{ij}\vert \leq \bar\epsilon_{ij}\Big\} \geq 1 - \bar\beta_{ij}, ~\text{with}~ \bar\epsilon_{ij}, \bar\beta_{ij} \in (0,1],
\end{align}
where $p_{ij}$ are entries of the probability transition matrix $\mathsf{\hat T}_{\mathsf x}$ that we are interested in its construction, defined as $p_{ij} = \mathsf{T}_{\mathsf x} (\Xi(x_j')\,\big|\,\bar x_i,u)$,
and $\bar p_{ij}$ are their empirical approximations as, $\forall \bar x_i \in \hat X, \forall u\in U$:
\begin{align}\label{Eq:4}
	&\bar p_{ij} = \frac{1}{G_{ij}}\sum_{z=1}^{G_{ij}} \mathbf{1}(x_z'\in\mathsf X_{j} \,\big|\, \bar x_i,u),\\\notag
	&\text{with}~~ \mathbf{1}(x_z'\in\mathsf X_{j} \,\big|\, \bar x_i,u)= \left\{\hspace{-1.7mm}
	\begin{array}{ll}
		1, \qquad\text{if} ~ x_z'\in\mathsf X_{j},\\
		0, \qquad\text{otherwise},
	\end{array}
	\right.
\end{align}
where  $G_{ij}$ is the required number of data for this empirical approximation and $x_z'$ is the state of the system after one-step transition from a current state $\bar x_i$ under an input $u$.
By a-priori fixing the threshold $\bar\epsilon_{ij}$ and confidence $\bar\beta_{ij}$, one can employ Chebyshev's inequality~\cite{saw1984chebyshev} and formally quantify the required number of data $G_{ij}$ for the empirical approximation~\eqref{Eq:4} via the following lemma.
\begin{lemma}\label{Lemma:1}
	For an a-priori approximation error $\bar\epsilon_{ij}\in(0,1]$ and a desired confidence $\bar\beta_{ij}\in (0,1)$, one has
	\begin{align*}
		\PP\Big\{\vert p_{ij}-\bar p_{ij}\vert \leq \bar\epsilon_{ij}\Big\} \geq 1 - \bar\beta_{ij},
	\end{align*}
	when $G_{ij} \geq \frac{1}{4\bar\beta_{ij}\bar\epsilon_{ij}^2}$.
\end{lemma}

The proof is similar to that of Lemma~\ref{Lem:2} by considering a Bernoulli distribution for the indicator function $\mathbf{1}(\cdot)$.

\begin{remark}
	Note that by choosing a fine threshold $\bar\epsilon_{ij}$ in the Chernoff bound in~\eqref{Chernoff}, the number of samples $G_{ij}$ would be large enough such that all cells will be visited.
\end{remark}

The Chernoff bound in~\eqref{Chernoff} indicates that $p_{ij}$ lives in an interval within a threshold $\bar\epsilon_{ij}$  with a confidence of at least $1 - \bar\beta_{ij}$, \emph{i.e.,}
\begin{align*}
	\PP\Big\{\bar p_{ij} - \bar\epsilon_{ij} \leq p_{ij}\leq \bar p_{ij} + \bar\epsilon_{ij}\Big\} \geq 1 - \bar\beta_{ij}. 
\end{align*}
The constructed abstraction via the proposed empirical approximation in~\eqref{Eq:4} is called an interval Markov decision process (IMDP)~\cite{Lavaei_Survey}.

\begin{remark}
	Note that since the constructed IMDP has confidences $\bar\beta_{ij}$ on its entries, one needs to consider an overall confidence  $\beta_3 = \sum_{i,j}\bar\beta_{ij}$ over the constructed IMDP. Accordingly, this confidence should be also added to the confidences of Theorem~\ref{Thm:6}, \emph{i.e.,} $\beta = \beta_1 + \beta_2 + \beta_3$.  
\end{remark}	

Since the results of Theorem~\ref{Thm:4} and inequality~\eqref{epsilon-perturbed} are tailored to finite MDPs, we now quantify the distance between probabilities of satisfaction of finite MDPs and of finite IMDPs, denoted by $\hat\Psi_\text{I}$, as
\begin{align}
	\big|\PP\big\{\hat\Psi_\text{I}\vDash\varphi^\varepsilon\big\} - \PP\big\{\hat\Psi\vDash\varphi^\varepsilon\big\}\big| \le \rho,
\end{align}
where $\rho = 2\mathcal T  \max_i\sum_{j = 1}^{n_{\bar x}}\bar\epsilon_{ij}$~\cite{baier2008principles,LMCS2015}.
In the case of having uniform intervals in IMDP,  \emph{i.e.,} $\bar\epsilon_{ij} = \bar\epsilon$, one has $\rho = 2\mathcal T\bar\epsilon n_{\bar x}$. Accordingly, inequality~\eqref{epsilon-perturbed} can be rewritten as
\begin{align}\label{guarantee}
	\PP\big\{\hat\Psi_\text{I}\vDash\varphi^\varepsilon\big\} - (\delta + \rho) \le \PP\big\{\Psi\vDash\varphi\big\}.
\end{align}

{\bf Computational complexity.} In order to provide the guarantee as in~\eqref{guarantee}, one needs $n_{\bar x}G_{ij}$ data points with $G_{ij}$ satisfying $G_{ij} \geq \frac{1}{4\bar\beta_{ij}\bar\epsilon_{ij}^2}$ as in Lemma~\ref{Lemma:1} to construct an IMPD, and $N$ data points as in Theorem~\ref{Thm:6} for solving $\text{SCP}_\varsigma$ in~\eqref{SCP1} and constructing an SBF between $\hat\Psi$ and $\Psi$.	

\subsection{Data-Driven Finite MDPs via Maximum Likelihood Estimation}\label{MLE}

Here, we provide an alternative approach to directly construct finite MDPs from data based on maximum likelihood estimation (MLE) methods~\cite{myung2003tutorial}. In particular,
since the proposed IMDP may require large amounts of data to provide tight intervals $\bar\epsilon_{ij}$ together with reasonable confidences $\bar\beta_{ij}$, especially when the dimension of the unknown system is high, we employ here maximum likelihood estimation (MLE) methods~\cite{myung2003tutorial} to estimate parameters of the probability distribution given some observed data. For instance, if the stochasticity is additive with a Gaussian distribution, then its mean and standard deviation via the MLE method can be estimated as
\begin{align*}
	\hat \mu_{\hat N} = \frac{1}{\hat N}\sum_{j = 1}^{\hat N} \bar x_j, \quad \hat\sigma_{\hat N}^2 = \frac{1}{{\hat N}-1}\sum_{j = 1}^{\hat N} (\bar x_j - \hat\mu_{\hat N})^2,
\end{align*}
where $\bar x_j$ are representative points and $\hat \mu_{\hat N},\hat \sigma_{\hat N}$ are the \emph{empirical} mean and standard deviation given the number of observed data $\hat N$.
One can then utilize the estimated parameters of the probability distribution and construct a finite MDP using the proposed approach in Section~\ref{MDPs}. 
It is worth mentioning that one can employ the MLE method and estimate parameters of \emph{any arbitrary} probability distribution. The finite MDP can then be constructed following  the approach of Section~\ref{MDPs} via estimated parameters obtaining from MLE methods.

\begin{remark}
	Note that it is possible in general to provide asymptotic confidence bounds for MLE approaches using Fisher information~\cite{le2012asymptotic}. However, since the confidence interval is asymptotic and for the sake of space, we do not present it in our work. Instead, we pick the required number of data $\hat N$ large enough such that one can get an accurate approximation for parameters of probability distributions. In the case study, we show empirically that a finite MDP constructed from the MLE method is very close to its model-based version.
\end{remark}

\section{Case Study}\label{Sec:Case}

In order to show the applicability of our data-driven results to \emph{controller synthesis} problems for \emph{nonlinear} stochastic systems, we apply them to the following nonlinear jet engine compressor~\cite{anta2010sample}:
\begin{align}\notag
	x_1(k+1) &= x_1(k) + \tau\big (-x_2(k) - \frac{3}{2}x_1^2(k) - \frac{1}{2}x_1^3(k)\big )+ 0.01 \varsigma_1(k),\\\label{jet}
	x_2(k+1) &= x_2(k) +  \tau\big (x_1(k) - \nu(k)\big ) + 0.01 \varsigma_2(k),
\end{align}
where $x_1 = \Lambda - 1, x_2 = \tilde\Lambda - \hat\Lambda - 2$, with $\Lambda,\tilde\Lambda,\hat\Lambda$ being, respectively, the mass flow, the pressure rise, and a constant, $u \in U = \{-0.5,-0.45,-0.4,\dots, 0.4,0.45,0.5\}$ is the control input, and $\tau = 0.01$ is the sampling time.
We assume that the model is unknown. 
The main goal is to construct a finite MDP together with an SBF from data by solving $\text{SCP}_\varsigma$ in~\eqref{SCP1}. We then employ the data-driven finite MDP as a suitable substitute of original unknown system and synthesize policies maintaining states of the unknown jet engine compressor in a safe set $X = [-0.5,0.5]^2$ for finite time horizons  with some probability of satisfaction and a desirable confidence level.

We fix the structure of SBF as $\mathcal S(q,x,\hat x) = q_1(x_1 - \hat x_1)^2 + q_2(x_2 - \hat x_2)^2 + q_3$. We also fix the threshold $\varepsilon_1 = 0.04$ and the confidence $\beta_2 = 0.01$ a-priori. 
Now we need to compute $\mathscr L_{g}$ which is required for computing the minimum number of data. We construct matrix $P$ based on coefficients of SBF. 
By considering $q_1,q_2\in[-0.01 , 0.01]$, we compute an upper bound for $\lambda_{\max}(P)$ as
$\lambda_{\max}(P) \le 0.02$. We fix $\eta = 0.05$.
Then, we compute  $\mathscr L_{g} = 9.39$, and accordingly, $\varepsilon_2 = 1.81 \times 10^{-5}$.
Since the number of decision variables affects the minimum required number of data in~\eqref{EQ:12},  we also fix $\psi = 0.047$ a-priori to reduce the number of decision variables to $4$. Now we have all the required ingredients to compute $N$.
The minimum number of data required for solving $\text{SCP}_\varsigma$ in~\eqref{SCP1} is computed as $N = 553559$. According to Lemma~\ref{Lem:2}, we fix $\mu = 0.005$, $\beta_{1} = 0.01$ and compute $M = 783$. 
We now solve the $\text{SCP}_\varsigma$ in~\eqref{SCP1} with the acquired $N, M$. Coefficients of SBF together with the optimal objective value of $\text{SCP}_\varsigma$ are computed as
\begin{align*}
	\mathcal S(q,x,\hat x)= 0.01(x_1- \hat x_1)^2 + 0.01(x_2 - \hat x_2)^2 + 16,~\Upsilon_\varsigma^*=-0.041.
\end{align*}
Since $\Upsilon^*_\varsigma + \varepsilon_1 = -0.001 \leq 0$, according to Theorem~\ref{Thm:6}, the constructed $\mathcal S$ from data is an SBF between finite MDP $\hat\Psi$ and unknown jet engine $\Psi$ with $\psi = 0.047$ and with a confidence of at least $1-\beta_1 - \beta_2 = 98\%$. Then according to Theorem~\ref{Thm:4}, by taking the initial states of $\Psi$ and $\hat\Psi$ as $[-0.3;0.3]$, we guarantee that the distance between states of $\Psi$ and $\hat\Psi$ will not exceed $\varepsilon = 0.7$ during the time horizon $\mathcal T= 5$ with a probability of at least $90\%$ and a confidence of at least $98\%$. Note that one can provide similar tight closeness guarantees for longer time horizons by selecting a finer discretization parameter $\eta$, and accordingly a smaller $\psi$, but at the cost of paying more computational complexity while solving SCP$_\varsigma$ in~\eqref{SCP1}.

\begin{figure}[h!]
	\centering 
	\includegraphics[width=0.38\linewidth]{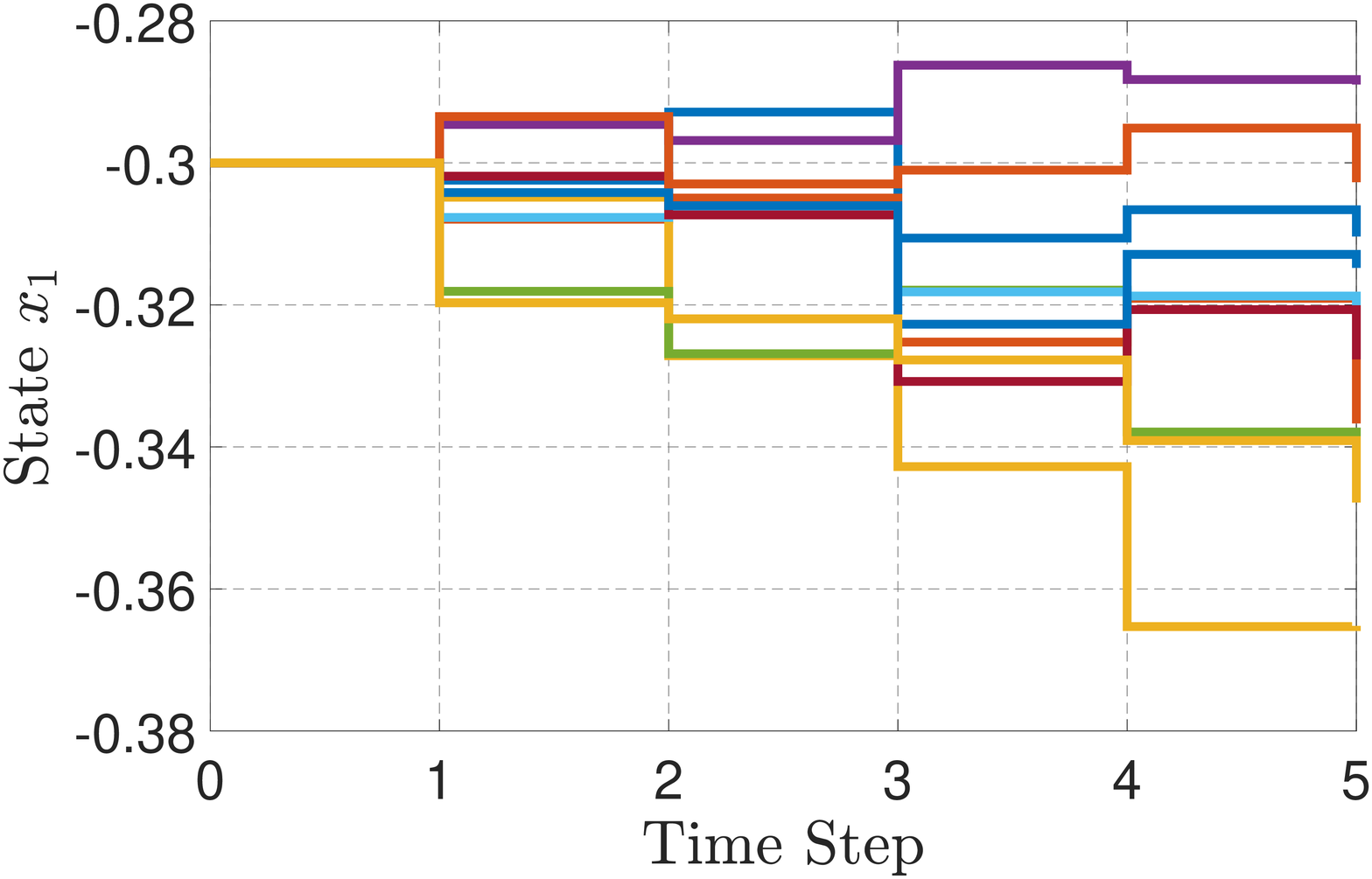}\hspace{0.4cm}
	\includegraphics[width=0.38\linewidth]{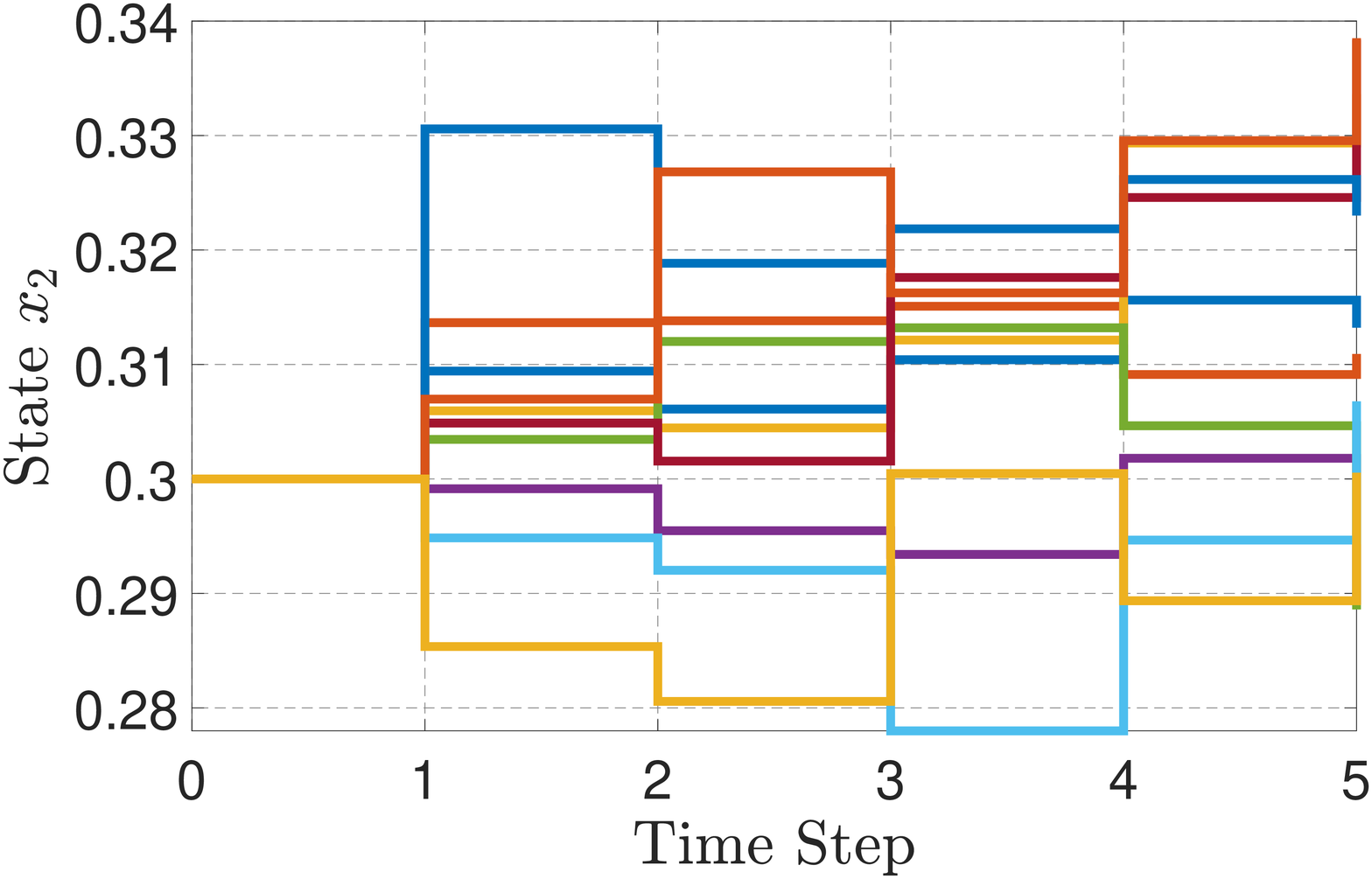}
	\includegraphics[width=0.38\linewidth]{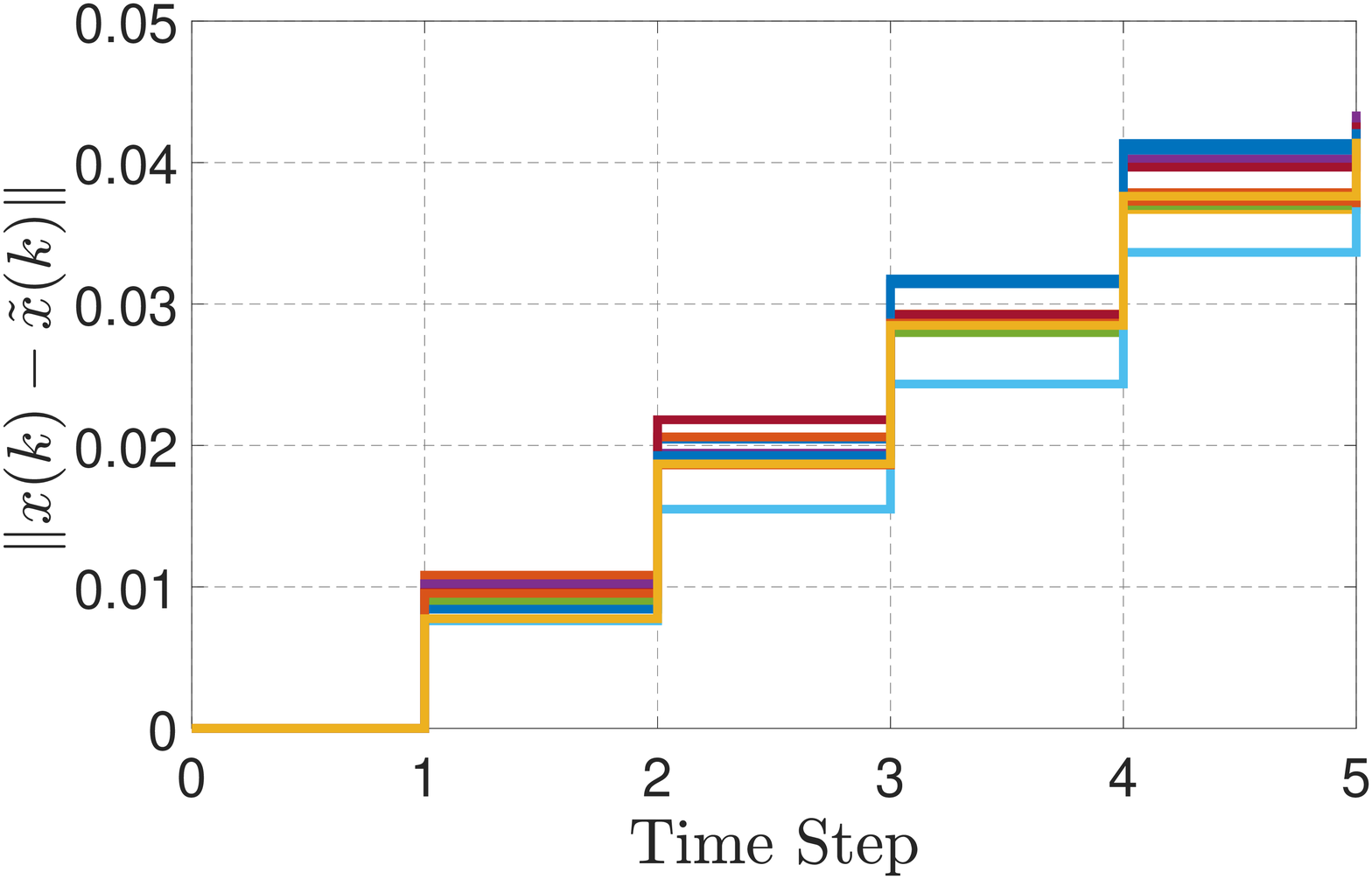}
	\caption{{\bf Top:} Closed-loop state  trajectories of unknown jet engine compressor with $10$ different noise realizations. As observed, the synthesized controller via the data-driven finite MDP keeps the trajectories of unknown jet engine in the comfort zone $[-0.5,~0.5]^2$. {\bf Bottom:} Norm of the error between state trajectories of $\Psi$, \emph{i.e.,} $\Vert x(k)-\tilde x(k)\Vert$, with the \emph{same noise realizations}, where $\tilde x$ is the closed-loop state trajectory of $\Psi$ generating from the \emph{model-based} finite MDP.}
	\label{Simulation}
\end{figure}

We now construct a finite MDP $\hat \Psi$ via the MLE method with $\hat N = 10^5$ and synthesize a controller for $\Psi$ via its data-driven finite MDP $\hat \Psi$ such that the controller maintains states of unknown jet engine in the comfort zone $[-0.5,0.5]^2$.
We employ the tool \texttt{AMYTISS} \cite{lavaei2020amytiss} to synthesize controllers for $\hat \Psi$. Note that although the domain of the controller is discrete, the synthesized policy for MDP is still state feedback (lookup table). Closed-loop state trajectories of the unknown jet engine compressor with $10$ different noise realizations are illustrated in Fig.~\ref{Simulation} top.

In order to show the quality of our data-driven finite MDP constructed from the MLE method, we now assume that we have access to the model of the jet engine compressor and construct a model-based finite MDP $\hat \Psi$. We synthesize controllers based on $\hat \Psi$ constructed from the model
via \texttt{AMYTISS} to keep states of the systems in the comfort zone $[-0.5,0.5]^2$.
In order to show that state trajectories of $\Psi$ generating from our data-driven approach are very close to their model-based version,
we now plot $\Vert x(k)-\tilde x(k)\Vert$ in Fig.~\ref{Simulation} bottom, with the \emph{same noise realizations}, where $\tilde x$ is the state trajectory of $\Psi$ generating from the \emph{model-based} finite MDP. As it can be observed, state trajectories of $\Psi$ based on the data-driven finite MDP constructed from MLE method are very close to their model-based version.

\section{Discussion}
	In this work, we proposed a data-driven technique for the construction of finite MDPs for discrete-time stochastic control systems with unknown dynamics. The main goal was to leverage stochastic bisimulation functions and quantify the probabilistic distance between unknown original systems and their data-driven finite MDPs while providing guaranteed error bounds. To do so, we cast the required conditions of SBF as a robust convex program (RCP). We then collected a finite number of data from trajectories of unknown systems and proposed a scenario convex program (SCP) associated to the original RCP. We established a probabilistic bridge between optimal values of the SCP and the RCP, and accordingly, formally constructed SBF between data-driven finite MDPs and unknown stochastic systems based on the number of data and a required confidence level. We verified our approaches by applying them to a \emph{nonlinear} jet engine compressor with unknown dynamics. Developing a \emph{compositional approach} for the \emph{data-driven} construction of finite MDPs for \emph{large-scale} stochastic systems and developing a more scalable IMDP based on the required number of data are under investigation as a future work.

\bibliographystyle{alpha}
\bibliography{biblio}

\section{Appendix}

\begin{lemma}\label{Lem:3}
	Consider a linear dt-SCS $x(k+1)=Ax(k) + B\nu(k)+ \varsigma(k)$ with $\varsigma(\cdot) \sim\mathcal N(0, \mathds{I}_n)$. Lipschitz constant $\mathscr{L}_{g}$ for a quadratic SBF of the form $(x-\hat x)^\top P(x-\hat x)$, with a positive-definite matrix $P\in\mathbb{R}^{n\times n}$, is computed as $\mathscr{L}_{g} = \max \big\{\mathscr{L}_{g_1},\mathscr{L}_{g_2}\big\}$, with
	\begin{align*}
		\mathscr{L}_{g_1}& = 2\lambda_{\max}(P) (2\mathscr L_1^2 h + 2\mathscr L_1 \mathscr L_2 \hat h + \mathscr L_1\eta + 2h),\\
		\mathscr{L}_{g_2}& = 4h (\lambda_{\min}(P) + \lambda_{\max}(P)),
	\end{align*}
	where
	$\Vert A\Vert \leq \mathscr L_1\in\mathbb R_0^+$, $\Vert B\Vert \leq \mathscr L_2\in\mathbb R_0^+$, $\Vert x\Vert \leq h\in\mathbb R_0^+$ for any $x\in X$, $\Vert u\Vert \leq \hat h\in\mathbb R_0^+$ for any $u\in U$.
\end{lemma}	

We now propose another  lemma for the computation of $\mathscr{L}_{g}$ for \emph{nonlinear} stochastic control systems.

\begin{lemma}\label{Lem:3_1}
	Consider a nonlinear dt-SCS $x(k+1)=f(x(k),\nu(k)) + \varsigma(k)$ with $\varsigma(\cdot) \sim\mathcal N(0, \mathds{I}_n)$. Lipschitz constant $\mathscr{L}_{g}$ for a quadratic SBF of the form $(x-\hat x)^\top P(x-\hat x)$, with a positive-definite matrix $P\in\mathbb{R}^{n\times n}$, is computed as $\mathscr{L}_{g} = \max \big\{\mathscr{L}_{g_1},\mathscr{L}_{g_2}\big\}$, with
	\begin{align*}
		\mathscr{L}_{g_1} &=2\lambda_{\max}(P) (2\mathscr L_f\mathscr L_x + \mathscr L_f\eta + 2h),\\
		\mathscr{L}_{g_2} &= 4h (\lambda_{\min}(P) + \lambda_{\max}(P)),
	\end{align*} 
	where $\Vert f(x,u)\Vert \leq \mathscr L_f\in\mathbb R_0^+ $, $\Vert \partial_{x}f(x,u) \Vert = \Vert \frac{\partial f(x,u)}{\partial x}\Vert \leq \mathscr L_{x}\in\mathbb R_0^+$, $\Vert x\Vert \leq h\in\mathbb R_0^+$ for any $x\in X$.
\end{lemma}

{\bf Proof of Lemma~\ref{Lem:2}.} By employing Chebyshev's inequality~\cite{saw1984chebyshev}, the difference between the expected value in $g_2$ and its empirical mean in $\bar g_2$ can be quantified as, $\forall x,x' \in X, \forall \hat x \in \hat X, \forall u \in U$,
\begin{align*}
	\PP\Big\{\Big|\EE \big  [\mathcal S(q,f(x,u,\varsigma),\hat{f}(\hat x,u,\varsigma))\,\big |\, x, \hat x,u \big ] -\frac{1}{M}\sum_{z=1}^{M}\mathcal{S}(q,f(x_i,u,\varsigma_{z}),\hat f(\hat x,u,\varsigma_{z}))\Big|\leq\mu\Big\}\geq 1-\frac{\sigma^2}{\mu^2},
\end{align*}
for any $\mu\in \mathbb{R}_0^+$, where
\begin{align*}
	&\sigma^2 = \text{Var}\Big[\frac{1}{M}\sum_{z=1}^{M}\mathcal{S}(q,f(x_i,u,\varsigma_{z}),\hat f(\hat x,u,\varsigma_{z}))\Big].
\end{align*}
Since $\text{Var}\big[\mathcal S(q,f(x,u,\varsigma),\hat{f}(\hat x,u,\varsigma))\big]\leq \mathcal{Q},\forall x \in X, \forall \hat x \in \hat X, \forall u \in U$, one has $\sigma^2 \leq \frac{\mathcal Q}{M}$. Accordingly, $\beta_{1}=\frac{\sigma^2}{\mu^2} \leq\frac{\mathcal{Q}}{M\mu^2}$. It implies that for $M \geq \frac{\mathcal{Q}}{\beta_{1}\mu^2}$: 
\begin{align*}
	\PP\Big\{\bar{\mathcal{S}}(q,x,\hat x) \models\text{SCP}_N\Big\}\geq 1- \beta_{1},
\end{align*} 
which completes the proof.$\hfill\blacksquare$

{\bf Proof of Theorem~\ref{Thm:6}.} We first establish a probabilistic
relation between optimal values of RCP and SCP$_N$. Based on~\cite[Theorem 4.1 and Remark 4.2]{esfahani2014performance}, the
probabilistic distance between optimal values of
RCP and SCP$_N$ can be formally lower bounded as
\begin{align}\label{EQ:12_1}
	\PP^N \Big\{0\leq\Upsilon^*_R-\Upsilon^*_N\leq\varepsilon_{1}\Big\}\geq 1-\beta_2,
\end{align}
provided that $$N\geq \bar N\big(w(\frac{\varepsilon_{1}}{\mathrm L_{\mathrm {SP}}\mathscr{L}_{g}}),\beta_2\big),$$ where
$w: [0,1]\rightarrow [0,1]$ is given by
\begin{align}\notag
	w(s) = s^{n}, \quad \forall s \in  [0,1],
\end{align}
and $\mathrm{L}_{\mathrm{SP}}$ is a Slater point as defined in~\cite[equation (5)]{esfahani2014performance}. Since the original RCP in~\eqref{RCP} is a $\min$-$\max$ optimization problem, the Slater constant  $\mathrm{L}_{\mathrm{SP}}$ can be selected as 1~\cite[Remark 3.5]{esfahani2014performance}. We refer the interested reader to ~\cite[equation (5)]{esfahani2014performance} for more details on the formal definition of Slater constant.

\noindent
From~\eqref{EQ:12_1}, one can readily conclude that $\Upsilon^*_N\leq\Upsilon^*_R\leq\Upsilon^*_N+\varepsilon_{1}$ with a confidence of at least $1-\beta_2$. From Lemma~\ref{Lem:2}, we have $\Upsilon^*_{N} \leq \Upsilon^*_{\varsigma}$ with a confidence of at least  $1-\beta_{1}$. Accordingly, one has $\Upsilon^*_{R}\leq\Upsilon^*_{N} + \varepsilon_{1} \leq \Upsilon^*_{\varsigma} + \varepsilon_{1}$. Since $\Upsilon^*_{\varsigma} + \varepsilon_{1} \leq 0$ (as the main condition of the theorem), it implies that $\Upsilon^*_{R} \leq 0$. Let us now define events $\mathcal E_1 :=\{ \Upsilon^*_{N} \leq \Upsilon^*_{\varsigma}\}$ and $\mathcal E_2 :=\{\Upsilon^*_{R}\leq\Upsilon^*_{N} + \varepsilon_{1}\}$, where $\PP\big\{\mathcal E_1\big\}\geq1-\beta_{1}$ and $\PP\big\{\mathcal E_2\big\}\geq1-\beta_2$. We are now computing the concurrent occurrence of  events $\mathcal E_1$ and $\mathcal E_2$ as:
\begin{align}\label{EQ:19}
	&\PP\big\{\mathcal E_1\cap \mathcal E_2\big\}=1-\PP\big\{\bar {\mathcal E_1}\cup \bar {\mathcal E_2}\big\},
\end{align}
where $\bar {\mathcal E_1}$ and $\bar {\mathcal E_2}$ are the complement of $\mathcal E_1$ and $\mathcal E_2$, respectively. Since
\begin{align*}
	&\PP\big\{\bar {\mathcal E_1}\cup \bar {\mathcal E_2}\big\}\leq\PP\big\{\bar {\mathcal E_1}\big\}+\PP\big\{\bar {\mathcal E_2}\big\},
\end{align*}
and by leveraging \eqref{EQ:19}, one can readily conclude that 
\begin{align}
	\PP\big\{\mathcal E_1\cap \mathcal E_2\big\}\geq 1-\PP\big\{\bar {\mathcal E_1}\big\}-\PP\big\{\bar {\mathcal E_2}\big\}\nonumber
	\geq 1-\beta_1 - \beta_{2}.
\end{align}
Then the constructed $\mathcal S$ from $\text{SCP}_\varsigma$ in~\eqref{SCP1} is an SBF between $\hat\Psi$ and $\Psi$ with a confidence of at least $1-\beta_1 - \beta_2$, which completes the proof.$\hfill\blacksquare$

{\bf Proof of Lemma~\ref{Lem:3}.} We first compute Lipschitz constants of $g_1, g_2$ with respect to $x$ and then take the maximum between them. For $g_2$, we have
\begin{align}\notag
	\mathscr{L}_{g_2}=\max\limits_{x\in X, \Vert x\Vert \leq h}\Vert\frac{\partial g_2}{\partial x}\Vert.
\end{align}
Accordingly,
\begin{align*} 
	&\mathscr{L}_{g_2} = \max\limits_{x\in X, \Vert x\Vert \leq h}\Vert 2((Ax + Bu) - \Pi_x(A\hat x+ Bu))^\top PA - 2(x-\hat x)^\top P\Vert\\
	&\leq \max\limits_{x\in X, \Vert x\Vert \leq h}\Vert  2((Ax + Bu) - \Pi_x(A\hat x+ Bu))^\top PA\Vert +\Vert 2(x-\hat x)^\top P\Vert\\
	& \leq  \max\limits_{x\in X, \Vert x\Vert \leq h} 2\Vert P \Vert\big(\Vert A \Vert(\Vert A x\Vert + \Vert Bu \Vert + \Vert \Pi_x(A\hat x + Bu)\Vert))+(\Vert x \Vert + \Vert \hat x \Vert)\big)\\
	& \leq  \max\limits_{x\in X, \Vert x\Vert \leq h} 2\Vert P \Vert\big(\Vert A \Vert(\Vert A x\Vert + \Vert Bu \Vert + \eta + (\Vert A\hat x+ Bu \Vert))+(\Vert x \Vert + \Vert \hat x \Vert)\big)\\
	& \leq 2\lambda_{\max}(P) (2\mathscr L_1^2 h + 2\mathscr L_1 \mathscr L_2 \hat h + \mathscr L_1\eta + 2h).
\end{align*}
For computing $\mathscr{L}_{g_1}$, since $\lambda_{\min}(P)\Vert x-\hat x \Vert^2 \leq (x-\hat x)^\top P(x-\hat x)$, we have $\alpha = \lambda_{\min}(P)$ in~\eqref{EQ:11}. Then one has
\begin{align*} 
	\mathscr{L}_{g_1} &=\max\limits_{x\in X, \Vert x\Vert \leq h}\Vert 2\lambda_{\min}(P) (x - \hat x) - 2(x-\hat x)^\top P\Vert\\
	& \leq 4h (\lambda_{\min}(P) + \lambda_{\max}(P)).
\end{align*}
Then $\mathscr{L}_{g} = \max \big\{\mathscr{L}_{g_1},\mathscr{L}_{g_2}\big\} = \max\big\{2\lambda_{\max}(P)(2\mathscr L_1^2 h + 2\mathscr L_1 \mathscr L_2 \hat h + \mathscr L_1\eta + 2h),4h (\lambda_{\min}(P) + \lambda_{\max}(P))\big\}$, which completes the proof.$\hfill\blacksquare$

{\bf Proof of Lemma~\ref{Lem:3_1}.} For $g_2$, we have
\begin{align*}
	&\mathscr{L}_{g_2} = \max\limits_{x\in X, \Vert x\Vert \leq h}\Vert  2(f(x,u) - \Pi_x(f(\hat x,u))^\top P\partial_{x}f(x,u)- 2(x-\hat x)^\top P\Vert\\
	& \leq  \max\limits_{x\in X, \Vert x\Vert \leq h} 2\Vert P \Vert\big(\Vert\partial_{x}f(x,u)\Vert(\Vert f(x,u)\Vert + \eta + \Vert f(\hat x,u)\Vert) + \Vert x \Vert + \Vert \hat x \Vert\big)\\
	& \leq 2\lambda_{\max}(P) (2\mathscr L_f\mathscr L_x + \mathscr L_f\eta + 2h).
\end{align*}
For $g_1$:
\begin{align*} 
	\mathscr{L}_{g_1} &=\max\limits_{x\in X, \Vert x\Vert \leq h}\Vert 2\lambda_{\min}(P) (x - \hat x) - 2(x-\hat x)^\top P\Vert\\
	& \leq 4h (\lambda_{\min}(P) + \lambda_{\max}(P)).
\end{align*}
Then $\mathscr{L}_{g} = \max \big\{\mathscr{L}_{g_1},\mathscr{L}_{g_2}\big\} = \max\big\{2\lambda_{\max}(P) (2\mathscr L_f\mathscr L_x + \mathscr L_f\eta + 2h),4h (\lambda_{\min}(P) + \lambda_{\max}(P))\big\}$, which completes the proof.$\hfill\blacksquare$

\end{document}